\documentclass[a4paper,10pt,two column]{revtex4}
\usepackage{graphicx} 
\usepackage{amsmath} 
\usepackage{amssymb} 
\usepackage{bm} 
\usepackage{dcolumn}
\usepackage{color}
\usepackage{mathrsfs}
\usepackage{amsfonts}
\usepackage{varioref}
\usepackage{mathrsfs}
\usepackage{graphicx}
\usepackage{latexsym}
\usepackage{amsmath}
\usepackage{amssymb}
\usepackage{textcomp}
\usepackage{amsbsy}
\usepackage{graphics}
\usepackage{epstopdf}
\usepackage{color}
\usepackage[caption=false]{subfig}

\RequirePackage[colorlinks,citecolor=blue,urlcolor=magenta,linkcolor=blue]{hyperref}
\input epsf

\allowdisplaybreaks[4]
\begin{document}

\tolerance=5000

\title{Natural validation of the second law of thermodynamics in cosmology}

\author{Sergei~D.~Odintsov$^{1,2}$\,\thanks{odintsov@ieec.uab.es},
Tanmoy~Paul$^{3}$\,\thanks{tanmoy.paul@visva-bharati.ac.in},
Soumitra~SenGupta$^{4}$\,\thanks{tpssg@iacs.res.in}} \affiliation{
$^{1)}$ ICREA, Passeig Luis Companys, 23, 08010 Barcelona, Spain\\
$^{2)}$ Institute of Space Sciences (ICE, CSIC) C. Can Magrans s/n, 08193 Barcelona, Spain\\
$^{3)}$ Department of Physics, Visva-Bharati University, Santiniketan 731235, India\\
$^{4)}$ School of Physical Sciences, Indian Association for the Cultivation of Science, Kolkata-700032, India}


\tolerance=5000

\begin{abstract}
The present work shows that the second law of thermodynamics gets naturally satisfied during the entire cosmic evolution of the universe starting from inflation to the late dark energy era, without imposing any exotic condition. This makes the inter-connection between cosmology and thermodynamics more concrete. Consequently, it also depicts that why the matter fields are not in thermal equilibrium with the apparent horizon during most of the cosmic era of the universe, except for the fluids with $\omega = -1/3$ leading to the transitions of the universe from an accelerating to a decelerating era and vice-versa.
\end{abstract}


\maketitle

The discovery of Bekenstein-Hawking entropy associated with an event horizon of a black hole interestingly brings two apparently  different arenas of physics, namely gravity and thermodynamics, on an equal footing \cite{PhysRevD.7.2333,Hawking:1975vcx,Bardeen:1973gs,Wald_2001}. One of the the distinctive features of Bekenstein-Hawking entropy of a black hole is that it depends on the area of the event horizon \cite{PhysRevD.7.2333,Hawking:1975vcx,Bardeen:1973gs,Wald_2001}, unlike  the classical thermodynamics where the entropy of a thermodynamic system depends on the volume of the same under consideration. Based on such an interesting feature of  Bekenstein-Hawking entropy, and depending on the non-additive statistics, various other form of entropies  have been proposed  such as the Tsallis \cite{Tsallis:1987eu} entropy, the R\'{e}nyi \cite{Renyi} entropy, the Barrow entropy \cite{Barrow:2020tzx}, the Sharma-Mittal entropy \cite{SayahianJahromi:2018irq}, the Kaniadakis entropy \cite{PhysRevE.72.036108}, or more generally, the generalized entropy \cite{Nojiri:2022dkr}. Regarding the black hole thermodynamics, few works have tried to relate the black hole thermodynamics with the Landauer principle \cite{Raju:2020smc,Herrera:2020dyh,Song_2007,KIM_2010,Bagchi:2024try}. Landauer principle sets the connection between thermodynamics and information theory, which states that a loss of one bit of information from a system to it's surrounding is associated with a dissipation of energy by  $\Delta Q \geq k_\mathrm{B}T\ln{2}$ where $k_\mathrm{B}$ is the Boltzmann constant and $T$ is the temperature of the surrounding in which the energy is dissipated \cite{5392446}. The equality sign in the Landauer principle holds for a reversible thermodynamical process where the system remains in thermal equilibrium to the surrounding; while the inequality sign points to irreversibility causes due to non-equilibrium between system and surrounding. Several recent works give strong support to the principle, both theoretically \cite{Bennett1973LogicalRO,Plenio_2001,Sagawa_2008} and experimentally \cite{Jun_2014,Berut:2012fei}. Very recently, \cite{Cortes:2024jvb} shows that the Hawking evaporation from a black hole satisfies the reversible case of the Landauer principle, leading to a firm connection between black hole thermodynamics and the information theory.

In the context of cosmology, the homogeneous and isotropic universe acquires an apparent horizon which, in analogy of black hole thermodynamics, may also be associated with an entropy; for the first studies of horizon related cosmology see for instance \cite{Jacobson_1995,Cai:2005ra,Cai:2006rs,Paranjape_2006,Komatsu:2015nkb,Mimoso:2016jwg} (for recent review of entropic cosmology, see \cite{Nojiri:2024zdu}). In the arena of entropic cosmology, the cosmological field equations are based on the first law of thermodynamics of the apparent horizon. However a consistent thermodynamic description of cosmology, also demands the validation of the second law of horizon thermodynamics, i.e., whether the change of total entropy (which is the sum of the horizon entropy and the entropy of the matter fields) proves to be positive with the cosmic expansion of the universe. In the present paper, we intend to do this. In particular, here we try to address the following questions:
\begin{itemize}
 \item Does the second law of thermodynamics get ``naturally'' satisfied during the cosmic evolution of the universe, without any exotic condition ?\\

 \item What about the thermal equilibrium between the apparent horizon and the matter fields inside the horizon ?
\end{itemize}
Moreover, we also examine that whether the validaton of second law of thermodynamics can be related to the Landauer principle in cosmological context.\\

The spatially flat Friedmann-Lema\^{i}tre-Robertson-Walker (FLRW) spacetime with the line element
\begin{eqnarray}
 ds^2 = -dt^2 + a^2(t)\left[dr^2 + r^2\left(d\theta^2 + \sin^2\theta d\varphi^2\right)\right] \, ,
 \label{metric}
\end{eqnarray}
admits an apparent horizon at the radius (with respect to comoving observer),
\begin{eqnarray}
 R_\mathrm{h} = \frac{1}{H} \, ,
 \label{app-hor}
\end{eqnarray}
where $a$ is the scale factor and $H = \dot{a}/a$ represents the Hubble parameter of the universe. The cosmic horizon is dynamical in nature, in particular $R_\mathrm{h}$ increases with time as long as the matter fields inside the horizon obey the null energy condition. Such dynamical nature of $R_\mathrm{h}$ plays a pivotal role in demonstrating the Landauer principle in cosmology. Another important quantity is the surface gravity on the apparent horizon: $\kappa = \frac{1}{2\sqrt{-h}}\partial_{a}\left(\sqrt{-h}h^{ab}\partial_{b}R\right)\big|_{R_\mathrm{h}}$ (with $h_{ab} = \mathrm{diag.}(-1,a^2)$ defines the induced metric along constant $\theta$ and constant $\varphi$, i.e. $h_{ab}$ is the induced metric along the normal of the apparent horizon) which, due to the spatially flat FLRW metric, is given by
\begin{eqnarray}
 \kappa = -\frac{1}{R_\mathrm{h}}\left(1 - \frac{\dot{R}_\mathrm{h}}{2}\right) \, .
 \label{surface gravity}
\end{eqnarray}
Below we briefly demonstrate that the apparent horizon may be associated with a thermal behaviour where the surface gravity defines the horizon temperature, namely,
\begin{eqnarray}
 T_\mathrm{h} = \frac{|\kappa|}{2\pi} = \frac{1}{2\pi R_\mathrm{h}}\left(1 - \frac{\dot{R}_\mathrm{h}}{2}\right) \, .
 \label{temperature}
\end{eqnarray}
In order to demonstrate the thermal behaviour of the cosmic horizon, let us consider the FLRW equation (in terms of $R_\mathrm{h}$):
\begin{eqnarray}
 H^2 = \frac{8\pi G}{3}~\rho \, ,
 \label{E-eqn}
\end{eqnarray}
where $\rho$ is the energy density of the matter fields inside the horizon. Eq.~(\ref{E-eqn}) has a differential form like,
\begin{eqnarray}
 -\left(\frac{2}{R_\mathrm{h}^3}\right)dR_\mathrm{h} = \frac{8\pi G}{3}~d\rho \, ,
 \label{diff-E-eqn}
\end{eqnarray}
which, owing to the conservation law of the matter fields: $\dot{\rho} + 3H\left(\rho + p\right) = 0$, takes the following form,
\begin{eqnarray}
 T_\mathrm{h}\left(\frac{2\pi R_\mathrm{h}}{G}\right)dR_\mathrm{h} = 4\pi R_\mathrm{h}^2\left(\rho + p\right)\left(1 - \frac{\dot{R}_\mathrm{h}}{2}\right)dt \, .
 \label{thermo-1}
\end{eqnarray}
The total internal energy stored in the matter fields inside the apparent horizon is given by $E = \rho V$, with $V = 4\pi/(3H^3)$ symbolizing the volume enclosed by the apparent horizon, and consequently, one has the following identity:
\begin{eqnarray}
 4\pi R_\mathrm{h}^2\left(\rho + p\right)dt = -dE + \rho dV \, ,
 \label{thermo-2}
\end{eqnarray}
where, once again, we have used the conservation of the matter fields to arrive at the above expression. Thereby Eq.~(\ref{thermo-1}) becomes,
\begin{eqnarray}
 T_\mathrm{h}\left(\frac{2\pi R_\mathrm{h}}{G}\right)dR_\mathrm{h} = \left(-dE + \rho dV\right)\left(1 - \frac{\dot{R}_\mathrm{h}}{2}\right) \, .
 \label{thermo-3}
\end{eqnarray}
Eq.~(\ref{thermo-3}) leads to two different formulations of thermodynamic law for the apparent horizon based on two different forms of the horizon entropy. In particular, the fist formulation comes as,
\begin{eqnarray}
 T_\mathrm{h}dS_\mathrm{h}^{(1)} = -dE + \frac{1}{2}\left(\rho - p\right)dV \, ,
 \label{thermo-4}
\end{eqnarray}
with
\begin{eqnarray}
 S_\mathrm{h}^{(1)} = \frac{2\pi}{G}\int R_\mathrm{h}dR_\mathrm{h} = \frac{A}{4G} \, ,
 \label{thermo-5}
\end{eqnarray}
while, the second formulation is given by,
\begin{eqnarray}
T_\mathrm{h}dS_\mathrm{h}^{(2)} = -dE + \rho dV \, ,
 \label{thermo-6}
\end{eqnarray}
with
\begin{eqnarray}
 S_\mathrm{h}^{(2)} = \frac{2\pi}{G}\int \frac{R_\mathrm{h}dR_\mathrm{h}}{\left(1 - \dot{R}_\mathrm{h}/2\right)} = \frac{A}{G\left(1 - 3\omega\right)} \, .
 \label{thermo-7}
\end{eqnarray}
Here $A = 4\pi R_\mathrm{h}^2$ represents the area of the horizon, and $\dot{R}_\mathrm{h} = 3(1+\omega)/2$ (with $\omega$ being the equation of state of the matter fields inside the horizon) due to the FLRW equations. Note that beside the form of horizon entropy, the work density term in the two thermodynamic formulations are different. However it is clear that $S_\mathrm{h}^{(2)}$ is ill-defined during the radiation dominated era when $\omega = 1/3$. This is the reason we will consider the first thermodynamic formulation for the apparent horizon, given by Eq.~(\ref{thermo-4}), in the rest of the paper where we express the horizon entropy by $S_\mathrm{h} = A/(4G)$ without any superfix. Thereby the horizon entropy corresponding to the FLRW equations of Einstein gravity is given by one-quarter of the area of the apparent horizon.

Owing to $S_\mathrm{h} \propto A = 4\pi/H^2$, the horizon entropy monotonically increases with cosmic time ($t$), in particular, we have
\begin{eqnarray}
 dS_\mathrm{h} > 0 \, ,
 \label{thermo-8}
\end{eqnarray}
provided the matter fields obey the null energy condition, i.e. $\omega > -1$. In the present work, we will not consider any phantom fields, in particular, the EoS of the matter fields satisfies $\omega > -1$.

\section*{Thermodynamics of matter fields}
\begin{figure}[!h]
\begin{center}
\centering
\includegraphics[width=2.0in,height=2.0in]{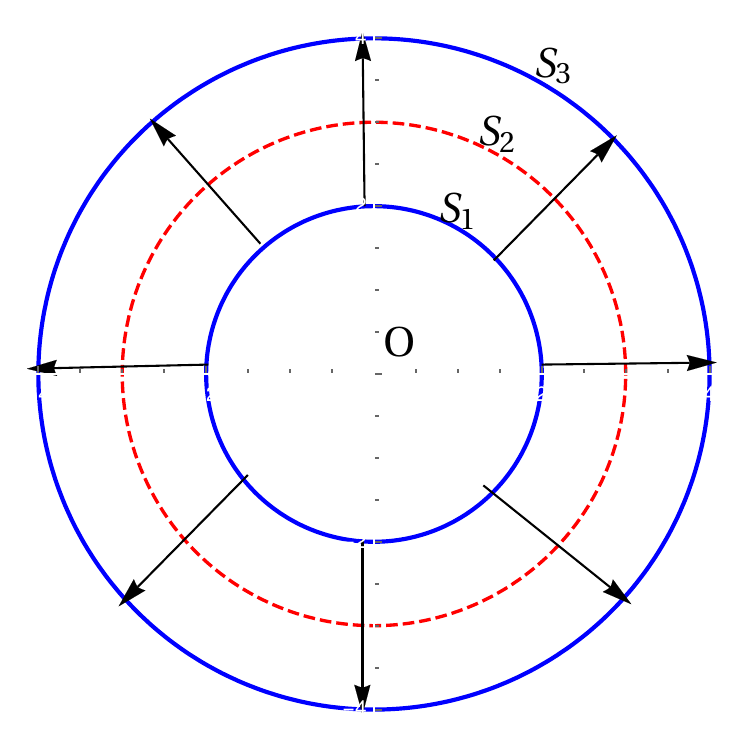}
\caption{Comparison between the formation of apparent horizon and the comoving expansion of the universe, in order to calculate the matter flux through the horizon.}
 \label{plot-2}
\end{center}
\end{figure}

We consider the matter fields inside the horizon to be a perfect fluid with a $constant$ equation of state (EoS) parameter $\omega$ given by:
\begin{eqnarray}
 p = \omega \rho \, ,
 \label{new-3}
\end{eqnarray}
where $p$ and $\rho$ represent the pressure and the energy density of the matter field, respectively. The above expression may be regarded as a barotropic equation of state, generally used in cosmology. Depending on the values of $\omega$, the universe undergoes through different cosmic stages. In the present work, we will use a general $\omega$ without putting any constraint on it. Our motive is to investigate that which values of the matter EoS parameter (naturally) allows the irreversibility or the reversibility of second law of thermodynamics in cosmological context.\\
The matter fields behaves like an open system as it exhibits a flux through the apparent horizon. Such kind of matter flux exists due to the the difference between the comoving expansion speed of the universe ($v_\mathrm{c}$) and the speed of the formation of the apparent horizon ($v_\mathrm{h}$). In particular, $v_\mathrm{c} = HD$ (at a physical distance $D$ from a comoving observer) and $v_\mathrm{h} = -\dot{H}/H^2$. Therefore the thermodynamic law of the matter fields inside the horizon can be expressed by,
\begin{eqnarray}
 T_\mathrm{m}dS_\mathrm{m}&=&(\mathrm{increase~of~internal~energy}) + (\mathrm{work~done})\nonumber\\
 &+&(\mathrm{energy~flux~through~horizon}) \, ,
 \label{thermo-m-1}
\end{eqnarray}
where $T_\mathrm{m}$ and $S_\mathrm{m}$ represent the temperature and entropy of the matter fields respectively. In general, the matter fields' temperature is considered to be different than the horizon temperature --- we will explicitly examine this issue at some stage. The internal energy of the matter fields (at instant $t$) is given by $E(t) = \rho(t)V(t)$ and thus the increase of internal energy during the cosmic interval $dt$ becomes (by using the conservation law of matter field),
\begin{eqnarray}
 dE = -3H\left(\rho + p\right)Vdt + \rho dV \, .
 \label{thermo-m-2}
\end{eqnarray}
Regarding the second term in Eq.~(\ref{thermo-m-1}), the work done by the matter fields is expressed as,
\begin{eqnarray}
 dW = \frac{1}{2}T_\mathrm{ab}h^{ab}dV \, ,
 \label{thermo-m-3}
\end{eqnarray}
here the work density is defined by the projection of the energy-momentum tensor of the matter fields along the normal of the apparent horizon \cite{Cai:2005ra,Cai:2006rs}, where $h^{ab}$ is the induced metric along the normal of the apparent horizon. Thus we have,
\begin{eqnarray}
\label{new-1}
d{s_\perp}^2 = \sum_{a,b=0,1} h_{ab} dx^a dx^b = - dt ^2 + a( t )^2 d r ^2 \, ,
\end{eqnarray}
i.e. $h_{ab} = \mathrm{diag.}(-1,a^2)$. Moreover, the energy-momentum tensor of the matter fields inside the horizon is given by: $T_\mathrm{\mu\nu} = \mathrm{diag.}(-\rho,p,p,p)$. As a consequence, one gets the following expression for the work done by the matter fields:
\begin{eqnarray}
 dW = \frac{1}{2}\left(p - \rho\right)dV \, ,
 \label{new-2}
\end{eqnarray}
For the third term in Eq.~(\ref{thermo-m-1}), we need to realize that the matter fields exhibit a flux through the horizon due to $v_\mathrm{c} \neq v_\mathrm{h}$, and the demonstration is shown in Fig.~[\ref{plot-2}] where the concentric spheres (with respect to the comoving observer labeled by 'O') represent as follows --- (a) $S_\mathrm{1}$: the visible universe bounded by the apparent horizon at time $t$, having radius $OS_\mathrm{1} = 1/H(t)$; (b) $S_\mathrm{2}$: the visible universe bounded by the apparent horizon at time $t+dt$, having radius $OS_\mathrm{2} = 1/H(t+dt) = \frac{1}{H}-\frac{\dot{H}}{H^2}dt$ (at the leading order in $dt$); (c) $S_\mathrm{3}$: due to the difference between $v_\mathrm{c}$ and $v_\mathrm{h}$, the sphere $S_\mathrm{1}$ moves from $S_\mathrm{1} \rightarrow S_\mathrm{3}$ due to the comoving expansion of the universe and thus the radius of $S_\mathrm{3}$ turns out to be $OS_\mathrm{3} = \frac{1}{H} + dt$ (as $v_\mathrm{c}(t) = 1$ on the apparent horizon).
Thereby we calculate,
\begin{eqnarray}
 V_\mathrm{c}(t+dt) - V(t+dt)&=&\frac{4\pi}{3}\left\{(OS_\mathrm{3})^3 - (OS_\mathrm{2})^3\right\}\nonumber\\
 &=&-\frac{2\pi}{H^2}\left(1 + 3\omega\right)dt \, ,
 \label{thermo-m-4}
\end{eqnarray}
depicting the gap between the comoving volume and the visible universe at time $t+dt$ (i.e. between the spheres $S_\mathrm{2}$ and $S_\mathrm{3}$). It is important to note that for $\omega > -1/3$ (i.e. during deceleration era of the universe), the comoving expansion speed is less than the speed of formation of the apparent horizon, which in turn leads to $V_\mathrm{c}(t+dt) < V(t+dt)$. On other hand, we get $v_\mathrm{c} > v_\mathrm{h}$ during accelerating universe with $-1 < \omega < -1/3$, resulting to the reverse scenario (recall that our regime of interest is $\omega > -1$). Thereby Eq.~(\ref{thermo-m-4}) immediately determines the outward flux of the matter fields' effective energy through the horizon as,
\begin{eqnarray}
 \mathrm{Flux}&=&\left(\rho + 3p\right)\times\left[V_\mathrm{c}(t+dt) - V(t+dt)\right]\nonumber\\
 &=&-\frac{2\pi\rho}{H^2}\left(1 + 3\omega\right)^2dt \, .
 \label{thermo-m-5}
\end{eqnarray}
In spirit of the matter flux, the factor $(\rho + 3p) = \rho(1+3\omega)$ can be realized as the chemical potential ($\mu$) of the matter fields as $\mu \propto -(\rho + 3p)$. It may be noted from Eq.~(\ref{thermo-m-4}) that $(1+3\omega)$ fixes the relative size between the spheres $S_\mathrm{2}$ and $S_\mathrm{3}$ (see Fig.~[\ref{plot-2}]). In particular, $(1+3\omega) > 0$ results to $V_\mathrm{c}(t+dt) < V(t+dt)$ depicting that the visible universe losses some matter particles during the cosmic expansion; while we get $V_\mathrm{c}(t+dt) > V(t+dt)$ for $(1+3\omega) < 0$, during when, the universe gains some matter particles inside the horizon. Therefore the direction of diffusion of matter particles from inside to outside the horizon is solely determined by the factor $(1+3\omega)$. Moreover $\mu \propto -(\rho + 3p)$ depicts that the chemical potential is negative for $\omega > -1/3$ while $\mu > 0$ for $-1 < \omega < -1/3$. This is expected as the universe with $\omega > -1/3$ undergoes through a deceleration era during when the matter fields experience an attractive gravitational force, that in turn results to $\mu < 0$; while the matter fields with $\omega < -1/3$ result to an accelerating universe and experience a kind of repulsive force that shows up through a positive valued chemical potential.

Owing to Eqs.~(\ref{thermo-m-2}), (\ref{thermo-m-3}) and (\ref{thermo-m-5}), the thermodynamic law for the matter fields inside the horizon from Eq.~(\ref{thermo-m-1}) turns out to be,
\begin{eqnarray}
 T_\mathrm{m}\frac{dS_\mathrm{m}}{dt} = -\frac{\pi\rho}{H^2}\left(3 + 10\omega + 15\omega^2\right) \, .
 \label{thermo-m-6}
\end{eqnarray}
This clearly shows that $T_\mathrm{m}\dot{S}_\mathrm{m} < 0$ in the range $\omega > -1$ of our interest, i.e. the entropy of the matter fields proves to monotonically decrease with the cosmic time.

From the previous discussions, particularly from Eq.~(\ref{thermo-8}) and Eq.~(\ref{thermo-m-6}), it becomes clear that the entropy of the horizon increases while the matter fields' entropy decreases with the time. Therefore we have,
\begin{eqnarray}
 T_\mathrm{h}\frac{dS_\mathrm{h}}{dt} > 0~~~~~\mathrm{and}~~~~~T_\mathrm{m}\frac{dS_\mathrm{m}}{dt} < 0 \, ,
 \label{exc-1}
\end{eqnarray}
which reveal that the heat energy is released by the matter fields and is absorbed by the apparent horizon. This immediately indicates that the flow of heat energy is directed from the matter fields to the apparent horizon during the cosmic expansion of the universe. Such spontaneous direction of heat flow in turn points the following inequality:
\begin{eqnarray}
 T_\mathrm{m} \geq T_\mathrm{h} \, .
 \label{exc-2}
\end{eqnarray}
This opens two different possibilities: $T_\mathrm{m} = T_\mathrm{h}$ or $T_\mathrm{m} > T_\mathrm{h}$. In the next two subsections we will examine which, out of these two possibilities, is allowed during the entire cosmic evolution of the universe started from inflation to the dark energy era.

Since the apparent horizon absorbs heat energy during the cosmic time, let us consider that the horizon entropy increases by an amount $\Delta S_\mathrm{h}$ within time $\Delta t$. During the increment of $\Delta S_\mathrm{h}$, the Hubble parameter changes by an amount (by using $S_\mathrm{h} = \pi/(GH^2)$):
\begin{eqnarray}
 \Delta H  = -\left(\frac{GH^3}{2\pi}\right)\Delta S_\mathrm{h} \, .
 \label{thermo-10}
\end{eqnarray}

\subsection*{Examination of Reversible ($T_\mathrm{m} = T_\mathrm{h}$) or Irreversible ($T_\mathrm{m} > T_\mathrm{h}$) cases of second law of thermodynamics}

Here we are going to examine whether the possibility $T_\mathrm{m} = T_\mathrm{h}$ is allowed by the cosmic evolution of the universe. In this case, the matter fields and the horizon should be in thermal equilibrium, and the heat flow from the matter fields $\rightarrow$ the horizon is reversible in nature. Therefore the total change of entropy should vanish, i.e.
\begin{eqnarray}
 \Delta S_\mathrm{h} + \Delta S_\mathrm{m} = 0 \, .
 \label{sub-1-1}
\end{eqnarray}
If $\left|\Delta Q_\mathrm{m}\right|$ is the amount of heat released by the matter fields (within the interval when the horizon entropy increases by $\Delta S_\mathrm{h}$), then
\begin{eqnarray}
 \left|\Delta Q_\mathrm{m}\right| = T_\mathrm{m}\left|\Delta S_\mathrm{m}\right| \, ,
 \label{sub-1-2}
\end{eqnarray}
which, due to Eq.~(\ref{sub-1-1}) along with $T_\mathrm{m} = T_\mathrm{h}$, takes the following form:
\begin{eqnarray}
 \left|\Delta Q_\mathrm{m}\right| = T_\mathrm{h}\Delta S_\mathrm{h} \, .
 \label{sub-1-3}
\end{eqnarray}
This is the reversible statement of second law of thermodynamics. By using Eq.~(\ref{thermo-m-6}), the above expression turns out to be,
\begin{eqnarray}
 \frac{\pi\rho}{H^2}\left(3 + 10\omega + 15\omega^2\right)\Delta t = T_\mathrm{h}\Delta S_\mathrm{h} \, ,
 \label{sub-1-4}
\end{eqnarray}
where $\Delta t$ is the time interval during when the horizon entropy increases by $\Delta S_\mathrm{h}$. Therefore, owing to Eq.~(\ref{thermo-10}), $\Delta t$ can be expressed by,
\begin{eqnarray}
 \Delta t = \frac{\Delta H}{\dot{H}} = -\left(\frac{GH^3}{2\pi\dot{H}}\right)\Delta S_\mathrm{h} \, .
 \label{sub-1-5}
\end{eqnarray}
Due to the above expression of $\Delta t$, along with Eq.~(\ref{temperature}) and $H^2 = 8\pi \rho G/3$, Eq.~(\ref{sub-1-4}) results to the following condition on $\omega$:
\begin{eqnarray}
 \frac{3 + 10\omega + 15\omega^2}{(1+\omega)(1-3\omega)} = 1 \, .
 \label{sub-1-6}
\end{eqnarray}
Therefore the reversible case of second law of thermodynamics demands the above condition to be hold for the matter EoS parameter. However Eq.~(\ref{sub-1-6}) holds true only for the fluids with $\omega = -1/3$, i.e. when the universe expands with no acceleration (or no deceleration). This clearly demonstrates that the reversible case of second law of thermodynamics is allowed at the transitions from an accelerating universe to a decelerating and vice-versa, for instance, the transitions from inflation to standard Big-Bang cosmology (SBBC) and from SBBC to the late dark energy era.\\

Let us now focus to examine the second possibility $T_\mathrm{m} > T_\mathrm{h}$. In this case, the matter fields are not in thermal equilibrium with the horizon and the heat exchange from the matter fields to the horizon should be irreversible in nature where
\begin{eqnarray}
 \Delta S_\mathrm{h} + \Delta S_\mathrm{m} > 0 \, .
 \label{sub-2-1}
\end{eqnarray}
Once again, if $\left|\Delta Q_\mathrm{m}\right| = T_\mathrm{m}\left|\Delta S_\mathrm{m}\right|$ is the amount of heat released by the matter fields (within the interval when the horizon entropy increases by $\Delta S_\mathrm{h}$), then we have the following inequality:
\begin{eqnarray}
 \left|\Delta Q_\mathrm{m}\right| > T_\mathrm{h}\Delta S_\mathrm{h}\left(\frac{\left|\Delta S_\mathrm{m}\right|}{\Delta S_\mathrm{h}}\right) \, ,
\label{sub-2-3}
\end{eqnarray}
where we use $T_\mathrm{m} > T_\mathrm{h}$. Due to Eq.~(\ref{sub-2-1}), the above expression gets automatically satisfied if,
\begin{eqnarray}
 \left|\Delta Q_\mathrm{m}\right| > T_\mathrm{h}\Delta S_\mathrm{h} \, .
 \label{sub-2-4}
\end{eqnarray}
This is the irreversible expression of second law of thermodynamics. By using Eq.~(\ref{thermo-m-6}), the above expression turns out to be,
\begin{eqnarray}
 \frac{\pi\rho}{H^2}\left(3 + 10\omega + 15\omega^2\right)\Delta t > T_\mathrm{h}\Delta S_\mathrm{h} \, .
 \label{sub-2-5}
\end{eqnarray}
We now can borrow $\Delta t$ from Eq.~(\ref{sub-1-5}) obtained in the previous subsection; by doing so, Eq.~(\ref{sub-2-5}) finally leads to the following condition on $\omega$:
\begin{eqnarray}
 \frac{3 + 10\omega + 15\omega^2}{(1+\omega)(1-3\omega)} > 1 \, .
 \label{sub-2-6}
\end{eqnarray}
Therefore the irreversible case of second law of thermodynamics demands the above condition to be hold on $\omega$. The interesting point to be noted that the condition (\ref{sub-2-6}) is well satisfied by all possible value of $-1 < \omega < 1/3$ except $\omega = -1/3$. Actually the situation with $\omega = -1/3$ satisfies the reversible case of second law of thermodynamics, as we showed in the previous subsection.

Here it deserves mentioning that depending on the  values of $\omega$, the universe undergoes through various stages during the cosmic evolution. In particular, $\omega > -\frac{1}{3}$ leads to a decelerated expansion of the universe, while $-1< \omega < -\frac{1}{3}$ results to an accelerated universe. During the Standard Big Bang Cosmology (SBBC), the universe is dominated by radiation (with $\omega = 1/3$) or by pressureless dust (with $\omega = 0$), leading to the usual decelerated cosmic expansion of the universe. On other hand, the early and the late phases of the universe undergoes through acceleration which demands some exotic matter fields with $-1 < \omega < -1/3$ violating the strong energy condition. Thus it turns out that the cosmic evolution of the universe, dominated by the fluids with $-1 < \omega < 1/3$, naturally satisfies the irreversible case of the second law of thermodynamics; except for the fluids having $\omega = -1/3$ leading to a no-acceleration (or no-deceleration) phase of the universe. Such irreversibility in cosmology is unlike to the black hole case where the Hawking evaporation admits the reversible scenario \cite{Cortes:2024jvb}. This points a crucial difference between black hole thermodynamics and the thermodynamics of cosmology.\\

Before concluding, we would like to mention that the second law of thermodynamics is connected with the Landauer principle which sets the connection between thermodynamics and information theory. In this regard, the minimum amount of information is stored within one bit; and thus the increase of the horizon entropy on gaining one bit of information becomes $\Delta S_\mathrm{h} = k_\mathrm{B}\ln{2}$ with the consideration that each bit of information has two degrees of freedom. Therefore, with $\Delta S_\mathrm{h} = k_\mathrm{B}\ln{2}$, if we perform the similar procedures as above, then the validity of the second law of thermodynamics can lead to the validity of the Landauer princple in cosmological context.\\

In conclusion, we show that the second law of thermodynamics gets $naturally$ satisfied in cosmology, without imposing any exotic condition. It turns out that the cosmic evolution of the universe with the matter fluids having $-1 < \omega < 1/3$ admits the irreversible scenario of second law of thermodynamics, except for the fluids having $\omega = -1/3$ leading to the transitions from an accelerating universe to a decelerating one and vice-versa. At such transition points (for instance --- the transitions from inflation to SBBC and from SBBC to the late time acceleration), the cosmic evolution obeys the reversible thermodynamic process. Therefore if we trace the cosmic era from inflation to the dark energy era, then the second law of thermodynamics shows as follows --- (a) Inflation and dark energy epochs with $-1 < \omega < -1/3$ allow irreversibility; (b) SBBC with $-1/3 < \omega < 1/3$ also obeys the irreversible case; (c) Transitions from inflation to SBBC, as well as from SBBC to dark energy, with $\omega = -1/3$ admits the reversible case. In this regard we would like to mention that a perfect fluid with $\omega < -\frac{1}{3}$, although produces an accelerating phase, may not result to a $viable$ inflation or a $viable$ dark energy era during the early and the late stages of the universe that are compatible with observational data. A viable inflation (or a viable dark energy era) typically requires modified theories of gravity, for instance some higher curvature gravity theory or non-minimally coupled scalar-tensor theory or by introducing a cosmological constant, where the cosmological field equations get modified compared to the Eq.~(\ref{E-eqn}). The validity of the Landauer principle in viable modified theories of gravity are expected to study in future. In the present work, we demonstrate that in Einstein cosmology with a perfect fluid inside the horizon (having a constant EoS parameter), how the irreversibility (or the reversibility) of the second law of thermodynamics gets naturally satisfied for different values of $\omega$ which are possibly connected with different cosmic stages of the universe. This inter-relates cosmology and thermodynamics in a more concrete way.

\bibliography{Landauerprcosmology-PRD}
\bibliographystyle{./utphys1}

\end{document}